\documentclass[twocolumn,showpacs,preprintnumbers,amsmath,amssymb]{revtex4}
\usepackage{psfig}
\usepackage{dcolumn}
\usepackage{bm}
\usepackage{float}
\begin{document}

\title{New results on the limit for the width of the  
exotic $\rm{\Theta^+}$ resonance}
\author{A. Sibirtsev} 
\affiliation{Institut f\"ur Kernphysik, Forschungszentrum 
J\"ulich, D-52425 J\"ulich, Germany \\
Special Research Centre for the 
Subatomic  Structure of Matter (CSSM) and Department of Physics and 
Mathematical Physics, University of Adelaide, SA 5005, Australia}
\author{J. Haidenbauer, S. Krewald}
\affiliation{Institut f\"ur Kernphysik, Forschungszentrum 
J\"ulich,  D-52425 J\"ulich, Germany}
\author{Ulf-G. Mei{\ss}ner}
\affiliation{Universit\"at Bonn, Helmholtz-Institut f\"ur Strahlen- 
und Kernphysik,  Nu{\ss}allee 14-16, D-53115 Bonn, Germany \\
Institut f\"ur Kernphysik, Forschungszentrum J\"ulich, D-52425 J\"ulich, 
Germany}
\preprint{FZJ-IKP(TH)-04/05}
\date{\today}
 
\begin{abstract}
We investigate the impact of the $\Theta^+$(1540) resonance on 
differential and integrated cross sections for the reaction $K^+d{\to}K^0pp$,
where experimental information is available at kaon momenta below 640~MeV/c.
The calculation utilizes the J\"ulich $KN$ model and extensions of it 
that include contributions from a $\Theta^+$(1540) state with different widths.
The evaluation of the reaction $K^+d{\to}K^0pp$ takes into account
effects due to the Fermi motion of the nucleons within the deuteron and the 
final three-body kinematics. We conclude that the available data constrain
the width of the $\Theta^+$(1540) to be less than 1~MeV.
\end{abstract}

\maketitle
Evidence for a narrow baryon resonance with positive strangeness, the $\Theta^+$,
has been found by more than ten experimental collaborations
with    masses ranging from $1521$ to $1555$~MeV 
and widths  from  $9$ and $24$~MeV (these widths are usually upper limits
given by the respective detector resolution)
\cite{review}.
The width of the $\Theta^+$ is of particular importance to understand the
nature of this state, see e.g.
\cite{Hong,Melikov}.
Constraints for the width of the $\Theta^+$ resonance can be deduced 
from 
$K^+N$ and $K^+d$ data which are available in  
the relevant momentum range, $417 < k_0 < 476\,$MeV/c, 
of the incident kaons 
\cite{Damerell,Stenger,Slater,Giacomelli1,Glasser,Giacomelli2}. 

A first, rough estimate for the width of the $\Theta^+$
based on $K^+d$ data was given by Nussinov~\cite{Nussinov}. Assuming
that there are fluctuations of about 2-4~mb in the experimental total
$K^+d$ cross section at momenta 500${\le}k_0{\le}$700~MeV that could be
due to the $\Theta^+$, he deduced an upper limit of $\Gamma_\Theta{<}$6~MeV.

A more refined estimate based on the $K^+d$ reaction was presented in
Ref.~\cite{Cahn}. Using data on the $K^+d{\to}K^0pp$ total cross
section for kaon momenta in the vicinity of the expected
$\Theta^+$ a conservative limit of 1~mb for the resonance cross
section was deduced and that led them to an upper limit for
the $\Theta^+$ width of around 1.1~MeV. A similar consideration
for the total $K^+d$ cross section resulted in a limit of
$\Gamma_\Theta$=0.8~MeV. Indeed, even for the certainly
unrealistic assumption that
the entire $I{=}0$ total cross section at $k_0$ = 440 MeV/c
is due to the $\Theta^+$ resonance, the width deduced by these authors for
the $\Theta^+$ did not exceed $\Gamma_\Theta$=3.6~MeV \cite{Cahn}.
A similar limit ($\Gamma_\Theta$=0.9~MeV) based on the total $K^+d$ cross 
section was also derived in Ref. \cite{Gibbs} using, however, only a 
selected set of data. 

The reexamination of the existing $KN$
data base in terms of a partial wave analysis, performed by 
Arndt, Strakovsky and Workman,  led  to the claim that 
widths of the $\Theta^+$ larger than a few MeV are excluded
~\cite{Arndt}. 
Specifically, it was found that the 
inclusion of a $\Theta^+$ resonance state with a width of 5~MeV 
in the $P_{01}$, $S_{01}$ or $P_{03}$ partial waves resulted in an 
increase of the total $\chi^2$ by 30\% or more. 

Similar conclusions were drawn from 
a direct comparison~\cite{Haidenbauer} between the available data
on total $KN$ cross sections in the $I{=}0$ and $I{=}1$ isospin channels
and a $KN$ model calculation based on the meson-exchange
model of the J\"ulich group \cite{Juel1,Juel2}. It was argued that the rather
strong enhancement of the cross section caused by the presence of a
$\Theta^+$ with a width of 20 MeV is not compatible with the
existing information on $KN$ scattering.
Only a much narrower $\Theta^+$ state, with a width in the order of 5 MeV or 
less, could be reconciled with
the existing data base, cf. Ref.~\cite{Haidenbauer} -- or, alternatively, the $\Theta^+$
state must lie at an energy much closer to the $KN$ threshold.

In this paper we want to extend the work of Ref.~\cite{Haidenbauer}.
We perform direct comparison of a calculation of the reaction $K^+d \to K^0pp$
with the corresponding experimental information.
One has to keep in mind that the information on the $K^+N$
interaction in the isospin $I=0$ channel has been inferred from
data on the $K^+d$ reaction. In the extraction procedure it is
implicitly assumed that the $K^+N$ amplitude shows no sharp structure.
Therefore, it is more conclusive to calculate explicitly observables for
the reactions $K^+d \to K^+np$ and $K^+d \to K^0pp$ based on
$K^+N$ interaction models that include a $\Theta^+$(1540) resonance
so that a direct comparison with experimental data is possible.
Then ``medium'' effects such as the broadening of the resonance
by the Fermi motion of the nucleons in the deuteron and the interaction of the
nucleons in the final state can be dealt with rigorously.

Starting point of the present investigation is again
the J\"ulich meson-exchange model for the $KN$
interaction. An extensive description of this model
is given in Refs.~\cite{Juel1,Juel2} where one can also find
its results for $KN$ phase shifts and for cross sections and polarizations.
Evidently this model yields a good overall reproduction of all
presently available empirical information on $KN$ scattering.
Specifically, it describes the data up to beam momenta of $k_0
\ \approx$ 1 GeV/c, i.e. well beyond the region of the observed
$\Theta^+$ resonance structure \cite{review}. Note that
the parameters of the model are fixed by a simultaneous fit to all
$KN$ partial waves and therefore the contributions to the $P_{01}$
channel, where the $\Theta^+$ pentaquark state is supposed to occur,
are constrained by the empirical information in the other partial waves.
We also utilize here the variants that were
presented in Ref. \cite{Haidenbauer}, where a $\Theta^+$ resonance was
added to the J\"ulich $KN$ model with a resonance position at 1540 MeV
and dynamically generated widths of 5 and 20 MeV, respectively, and we
consider
two more variants with widths of 1 and 10 MeV, constructed in the same way
as described in Ref. \cite{Haidenbauer}.

When adding a new ingredient, in the form of the $\Theta^+$
resonance, to the $KN$ model of the J\"ulich group one should,
in principle, refit all the free parameters of this model.
However, in practice it turned out that the available experimental
information (i) on those $KN$ partial-wave amplitudes where the
$\Theta^+$ does not contribute (i.e. all except the $P_{01}$)
and (ii) on the behaviour of the $P_{01}$ amplitude at higher energies,
i.e. away from the $\Theta^+$ resonance region, provides rather strong
constraints on the model parameters and therefore even a very moderate
change in those parameters would already lead to a deterioration of the
overall description of the $KN$ data.
Moreover, one has to keep in mind that the magnitude of the
$KN$ cross section generated by a resonance is determined by
unitarity constraints only, for energies below the inelastic
threshold \cite{Nussinov}. It cannot be changed by varying the parameters
of the model, anyway.
Let us also mention that the $\Theta^+$ resonance is added to the $KN$ model 
on the potential level, cf. Refs. \cite{Haidenbauer,Sibi} for details. In 
this way ambiguities with regard to the relative phase are avoided and the
interference pattern follows directly from the underlying dynamics as
discussed in Ref. \cite{Sibi}.

For the calculation of the reaction $K^+d \to K^0pp$
we follow, in general, the theoretical procedure
which was originally developed by Stenger et al.~\cite{Stenger}.
A detailed description of the formalism can be  found in
Ref.\cite{Hashimoto}.

The amplitude $T_d$ for the deuteron breakup reaction is 
\begin{eqnarray}
T_d =\sqrt{16\pi^3m_d}[T_N(q)u(p)+T_N(p)u(q)],
\label{eq1}
\end{eqnarray}
where $m_d$ is the deuteron mass, $p$ and $q$ are the momenta of 
the two final nucleons, $u$ is the ($S$-wave) deuteron wave function and
$T_N$ is elementary $KN$ amplitude.
In the present calculation we used the deuteron wave function of the
CD-Bonn potential. Exploratory calculations based on other wave functions
(Paris, Hulthen) indicated, however, that the results are rather insensitive 
to the specific choice. Note that throughout we neglect the deuteron D-state.
The differential cross section
for the $K^+d{\to}K^0pp$ reaction is then given 
as~\cite{Stenger,Hashimoto} 
\begin{eqnarray}
\frac{d\sigma}{d\Omega}{=}(|f_x|^2{+}|g_x|^2) [I(\theta){-}
J(\theta)]+\frac{2}{3}|g_x|^2J(\theta),
\label{eq2}
\end{eqnarray}
where $f_x$ and $g_x$ are the elementary spin-non-flip and spin-flip 
$K^+n{\to}K^0p$ amplitudes and $I$ and $J$ are the deuteron inelastic 
form factors, respectively. They are explicitely 
given by~\cite{Stenger,Hashimoto} 
\begin{eqnarray}
I&{=}&F\!\!\int\!\!\frac{k^2dk}{E_K}\frac{d^3p}{E_p}\frac{d^3q}{E_q}
\delta^4(k_0{+}P{-}k{-}p{-}q) \frac{u^2(p){+}u^2(q)}{2},\nonumber\\
\label{eq3}
J&{=}&F\!\!\int\!\!\frac{k^2dk}{E_K}\frac{d^3p}{E_p}\frac{d^3q}{E_q}
\delta^4(k_0{+}P{-}k{-}p{-}q)u(p)u(q),
\label{eq4}
\end{eqnarray}
where $k_0$, $k$ and $P$ are the momenta of the initial and final 
kaon and of the deuteron, respectively. $E_k$, $E_p$ and $E_q$
are the total energies of particles in the final state. 
The factor $F$ accounts for 
the transformation of the kaon scattering angle $\theta$ from 
the laboratory deuteron rest frame to the center-of-mass frame of 
the $KN$ two-body system. It is usually evaluated in the 
stationary spectator configuration, i.e. by assuming that the 
reaction takes place on the neutron at rest.

In the derivation of the expression for the $K^+d{\to}K^0pp$ 
differential cross section one encounters three-body phase 
space integrals of the $KN$ amplitudes over the 
momentum distribution of the nucleons within the deuteron. 
The form factor approximation rests on the 
assumption that the elementary amplitudes $f_x$ and $g_x$ and the
kinematic factors vary only slightly over the integration range 
and therefore can be taken out of the integrands and evaluated 
for a fixed typical nucleon momentum. The remaining integrals are 
then the form factors $I$ and $J$. 
However, in the presence of the $\Theta^+$ resonance 
the amplitudes $f_x$ and $g_x$ depend
strongly on the kaon energy and can not be removed from the $I$
and $J$ integrands. Therefore, in our analysis we integrate the $KN$
amplitude over the final three-body phase space. Furthermore we
do not use the stationary spectator approximation but compare our
calculations directly to the differential cross sections measured
in the deuteron rest frame.

\begin{figure}[hbt]
\vspace*{-5mm}
\hspace*{-2.mm}\psfig{file=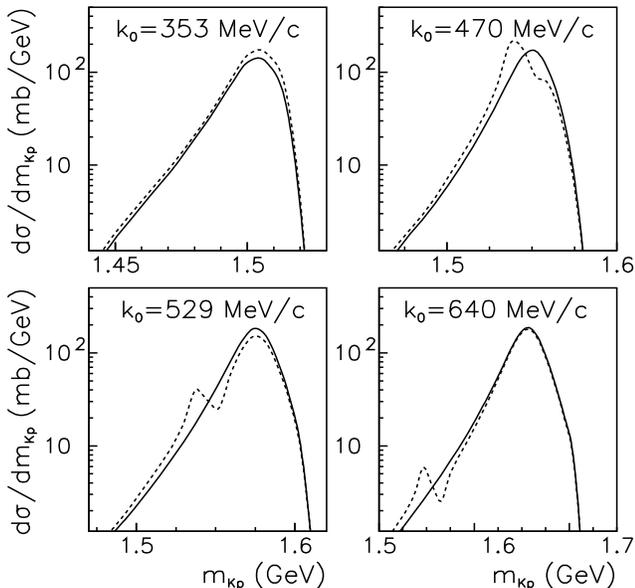,height=9.cm,width=9.3cm}
\vspace*{-11mm}
\caption{\label{penta4} The $K^0p$ invariant mass spectra from
$K^+d{\to}K^0pp$ reaction  at different $K^+$-meson momenta. Solid lines
show our calculations without a $\Theta^+$(1540) resonance, 
while the dashed lines 
indicate the results obtained with $\Gamma_\Theta$=5~MeV.} 
\end{figure}

It is worthwhile to mention that for the $K^+d{\to}K^0pp$ three-body 
final-state the invariant mass $m_{Kp}$ of the $K^0p$ system is integrated
over the range from $m_K{+}m_p$ to $\sqrt{s}{-}m_p$, where
$m_K$ and $m_p$ stand for the masses of the $K^0$-meson and the proton, 
respectively, and  
\begin{eqnarray}
s=m_K^2+m_d^2+2m_d\sqrt{m_k^2+k_0^2}.
\label{eq5}
\end{eqnarray}
Therefore, for a fixed initial kaon momentum, $k_0$, the
deuteron experiment samples the elementary $K^+n{\to}K^0p$
amplitude over the $m_{Kp}$ range given above. This situation 
substantially differs from the ``free neutron target'' approximation, 
where the  invariant mass $m_{Kp}$ is fixed by $k_0$ through 
Eq.~(\ref{eq5}). For the ``free target'' measurements the $\Theta^+$ mass of 
1530~MeV corresponds to incident kaon momentum of $k_0$=417~MeV/c and
only the data around that momentum can be sensitive to the 
$\Theta^+$ resonance. 
If one does not make the assumption of a free target,  
all $K^+d{\to}K^0pp$ observables above 
$k_0$=417~MeV/c will be influenced by the presence of the $\Theta^+$ 
resonance, which 
will show up in the $K^0p$ mass distribution. Note that the
$m_{Kp}$ spectrum is affected by the deuteron wave function, since the
maximal $K^0p$ mass corresponds to the minimal spectator momentum,
while the minimal $m_{Kp}$ probes high spectator momenta. In addition the
$\Theta^+$ resonance occupies only a small fraction of the $K^0p$ mass 
distribution, while the overall $m_{Kp}$ integration includes 
large part of the ``non-resonant background''. Therefore it might be that
the $\Theta^+$ signal in the invariant $K^0p$ mass spectrum becomes 
invisible after $m_{Kp}$ integration. 
The arguments given above are confirmed by our explicit calculations
of the $K^0p$ mass spectra for different kaon momenta, which are shown 
in Fig.~\ref{penta4}. Here the solid lines show the calculations
without $\Theta^+$ resonance, while the dashed lines are our results with
a $\Theta^+$(1540) with a width $\Gamma_\Theta$=5~MeV.

\begin{figure}[htb]
\vspace*{-7mm}
\hspace*{-4.mm}\psfig{file=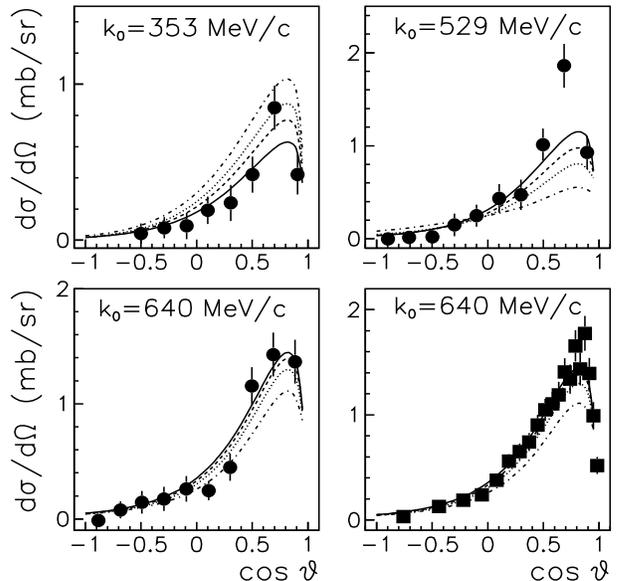,height=9.cm,width=9.5cm}
\vspace*{-12mm}
\caption{\label{penta1} The $K^0$-meson angular spectra from
the $K^+d{\to}K^0pp$ reaction for different $K^+$ momenta. 
The curves show our results for the original J\"ulich $KN$ 
model without the $\Theta^+$ resonance (solid line) and the variants with 
a $\Theta^+$(1540) and with different widths ($\Gamma_\Theta$=5~MeV - dashed; 
10~MeV - dotted; and 20~MeV - dash-dotted). 
The data are from Slater et al.~\cite{Slater} (circles) and
Giacomelli et al.~\cite{Giacomelli1} (squares). 
} 
\end{figure}
\begin{figure}[htb]
\vspace*{-5mm}
\hspace*{-3.mm}\psfig{file=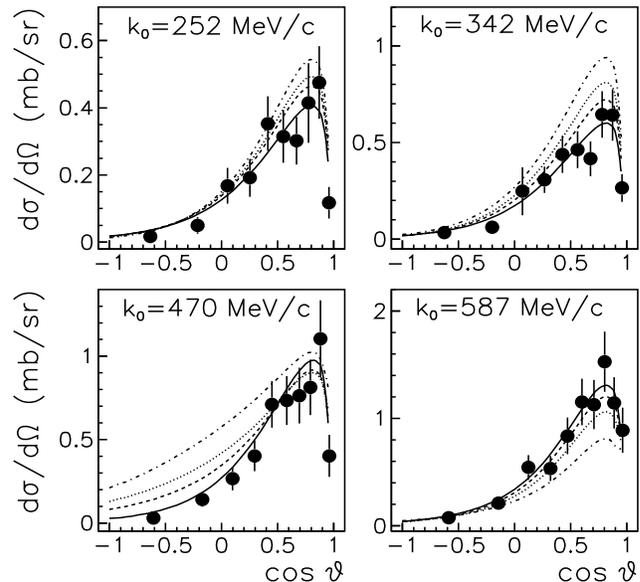,height=9.cm,width=9.5cm}
\vspace*{-12mm}
\caption{\label{penta2} The $K^0$-meson angular spectra from
$K^+d{\to}K^0pp$ reaction at different $K^+$ momenta. 
For notations, see Fig.~\ref{penta1}. 
The data are from Glasser et al.~\cite{Glasser}. 
} 
\end{figure}
\begin{figure}[htb]
\vspace*{-5mm}
\hspace*{-3.mm}\psfig{file=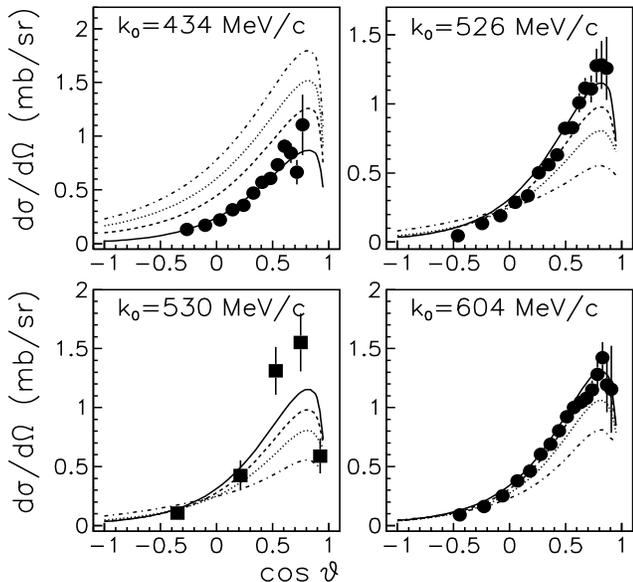,height=9.cm,width=9.5cm}
\vspace*{-12mm}
\caption{\label{penta5} The $K^0$-meson angular spectra from
$K^+d{\to}K^0pp$ reaction at different $K^+$ momenta. 
For notations, see Fig.~\ref{penta1}.
The data from Damerell et al.~\cite{Damerell} (circles)
and Stenger et al.~\cite{Stenger} (squares). 
} 
\end{figure}

The $K^0$-meson angular spectra for the reaction
$K^+d{\to}K^0pp$ at different $K^+$-meson momenta are
shown in Figs.~\ref{penta1}, \ref{penta2}, and \ref{penta5}. The 
lines correspond to calculations with the original J\"ulich $KN$ model
(i.e. without a $\Theta^+$) and with the variants with different 
$\Theta^+$ widths.  Note that the resonance energy is always the
same and was assumed to be 1540 MeV. The spin and parity is assigned
as $1/2^+$.
Our results are compared with the available experimental 
spectra~\cite{Slater,Giacomelli1,Glasser,Damerell} that have 
kaon momenta below 640~MeV/c.
It is interesting to see that 
the calculations with a $\Theta^+$ increase the 
differential $K^+d{\to}K^0pp$ cross sections at momenta 
$k_0{\le}$470~MeV/c and decrease them at higher momenta
as compared to the those obtained without a $\Theta^+$ resonance.
This effect is caused by the interference of the $\Theta^+$ and
the non-resonant $P_{01}$ contribution and is clearly illustrated 
by the $K^0p$ invariant mass distribution shown in 
Fig.~\ref{penta4}. Note also that at large 
kaon momenta the integration over the $K^0p$ invariant mass 
does not allow to distinguish between the situation with 
$\Gamma_\Theta{\le}$10~MeV and that without a $\Theta^+$ resonance.
 
A detailed inspection 
of the available differential $K^+d{\to}K^0pp$ cross sections
clearly indicates that the measurement by Glasser et 
al.~\cite{Glasser} at the $K^+$-meson momentum $k_0$=470~MeV/c
and of Damerell et al.~\cite{Damerell} at $k_0$=434~MeV/c 
are the most crucial ones for the determination of the $\Theta^+$ 
width. 
By comparing our results with the 138 experimental points on
differential $K^+d{\to}K^0pp$ cross sections shown in 
Figs.~\ref{penta1}, \ref{penta2} and \ref{penta5} we can deduce a $\chi^2$/dof.
Table~1 lists the $\chi^2$/dof evaluated for different 
$\Theta^+$ widths. To emphasize the impact of the spectra 
measured at $k_0$=434~MeV/c~\cite{Damerell} and
$k_0$=470~MeV/c~\cite{Glasser} we also present the 
$\chi^2$ obtained by excluding them from the analysis, which 
is indicated as solution $B$ in Table~1. 

\begin{table}[htb]
\caption{\label{tab1} $\chi^2$/dof evaluated by comparing
our calculations for different $\Theta^+$ widths, $\Gamma_\Theta$, 
with the experimental information on $K^0$ angular 
spectra measured in the reaction $K^+d{\to}K^0pp$ at kaon momenta from
252 to 640 MeV/c (138 data points). 
Result $A$ was obtained by analyzing the data
shown in Figs.\ref{penta1},\ref{penta2} and \ref{penta5}. 
Result $B$ was obtained 
by excluding the $K^0$ spectra at $k_0$=434 MeV/c~\cite{Damerell} 
and $k_0$=470 MeV/c~\cite{Glasser}.}
\begin{ruledtabular}
\begin{tabular}{lcccc}
$\Gamma_\Theta$ (MeV) & 0 & 5 & 10 & 20\\
\hline
$A$ & 1.8 & 7.4 & 27.7 & 42.2 \\
$B$ & 1.4 & 1.7 & 2.4 & 4.8 \\
\end{tabular}
\end{ruledtabular}
\end{table}

\begin{figure}[t]
\vspace*{-4mm}
\hspace*{-3.mm}\psfig{file=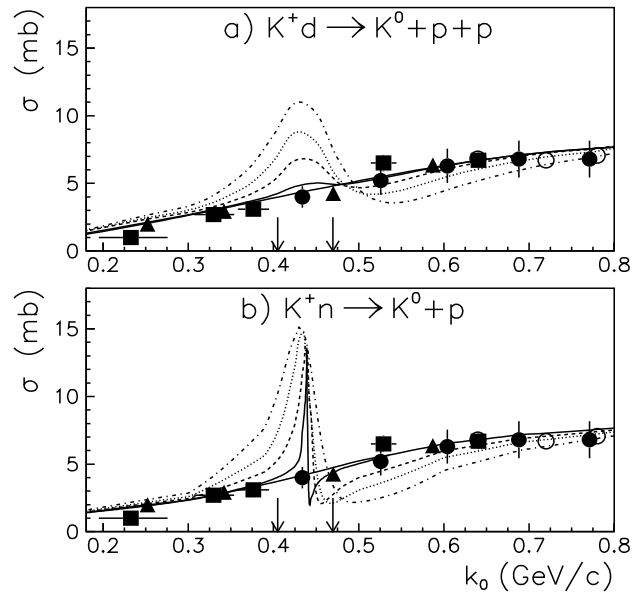,height=9.cm,width=9.3cm}
\vspace*{-10mm}
\caption{\label{penta3} Total $K^+d{\to}K^0pp$ cross 
section as a function of the kaon momentum. 
The curves in a) show our full results for the original J\"ulich $KN$ 
model without the $\Theta^+$ resonance (solid line) and the variants with 
a $\Theta^+$ and with different widths ($\Gamma_\Theta$=1~MeV - solid with bump; 
5~MeV - dashed; 
10~MeV - dotted; and 20~MeV - dash-dotted). 
The curves in b) correspond to a calculation for the reaction
$K^+n{\to}K^0p$ assuming that the neutron target is at rest.
Data are from Refs. \cite{Damerell} (filled circles),
Ref.~\cite{Slater} (squares), \cite{Glasser} (triangles)
and \cite{Giacomelli1,Giacomelli2} (open circles).
The vertical arrows indicate the range of kaon momenta corresponding
to the smallest and the largest values found experimentally for the
mass of the $\Theta^+$ resonance.
} 
\end{figure}

In Fig.~\ref{penta3}a we present results for the integrated $K^+d{\to}K^0pp$ 
cross section as a function of the kaon momentum in comparison to the
available experimental information. Again, 
it is evident that the data of Glasser et al. \cite{Glasser}
and Damerell et al. \cite{Damerell}, as mentioned above, 
provide the most restrictive constraints for the $\Theta^+$ width. 
Thus, it is certainly fortunate that there are two 
independent measurements in the critical energy range.
One can see from Fig.~\ref{penta3}a that none of the model calculations with 
a $\Theta^+$ width larger than 1 MeV is compatible with the data. Widths of
1 MeV or less can be certainly accommodated though we should say here that we
did not explore the effect of such narrow widths in an actual model calculation. 
 
If we disregard again the two data points from Refs. \cite{Glasser}
and \cite{Damerell}, respectively, there is a larger gap in the data base
just at those energies where the $\Theta^+$ is supposed to be located 
(the largest and smallest resonance masses reported so far are indicated by
bars in Fig.~\ref{penta3}) and that allows to fit in such a resonance with a width 
of $\Gamma_\Theta \approx$ 5 MeV without increasing the $\chi^2$/dof by more
than 10\%, cf. Table 1. 

In order to illustrate the impact of the stationary 
neutron approximation we show here also calculations for the
two-body reaction $K^+n{\to}K^0p$, cf. Fig.~\ref{penta3}b. 
Comparing the two panels of the figure one can see to which extend
the resonance is broadened by the Fermi motion of the nucleons in 
the deuteron and by the integration over the three-body phase space. 

In summary, we have investigated the impact of the $\Theta^+$(1540) resonance 
on the reaction $K^+d{\to}K^0pp$
where experimental information is available at kaon momenta below 640~MeV/c.
The calculation utilizes the J\"ulich $KN$ model and extensions of it 
that include contributions from a $\Theta^+$(1540) state with different widths.
The evaluation of the reaction $K^+d{\to}K^0pp$ takes into account
effects due to Fermi motion of the nucleons within the deuteron and the 
final three-body kinematics. The comparison with existing data on 
differential and integrated cross sections suggests that there is no
room for a $\Theta^+$ resonance with a width of more than 1~MeV. 

\medskip\noindent
We appreciate  discussions with W. Eyrich, C. Hanhart, 
A.W. Thomas and I. Zahed. 
This work was partially supported by the grant 
N.~447AUS113/14/0 by the Deutsche Forschungsgemeinschaft and the 
Australian Research Council.

\end{document}